\documentclass[preprint,showpacs,preprintnumbers,amsmath,amssymb,prb]{revtex4}
\usepackage{graphicx}% Include figure files
\usepackage{dcolumn}% Align table columns on decimal point
\usepackage{bm}% bold math
\usepackage{epsfig}
\begin{document}

\preprint{PREPRINT}

\title{Structural anomalies for a three
dimensional isotropic core-softened potential}

\author{ Alan Barros de Oliveira}
\affiliation{ Instituto de F\'{\i}sica, Universidade Federal do Rio 
Grande do Sul,  
Caixa Postal 15051, 91501-970, Porto Alegre, RS, BRAZIL.}

\author{Paulo A. Netz}
\affiliation{ Instituto de Qu\'{\i}mica, Universidade Federal do Rio 
Grande do Sul,  
91501-970, Porto Alegre, RS, BRAZIL.}

\author{ Thiago Colla}
\affiliation{ Instituto de F\'{\i}sica, Universidade Federal do Rio 
Grande do Sul, 
Caixa Postal 15051, 91501-970, Porto Alegre, RS, BRAZIL}

\author{ Marcia C. Barbosa}
\affiliation{ Instituto de F\'{\i}sica, Universidade Federal do Rio 
Grande do Sul, 
Caixa Postal 15051, 91501-970, Porto Alegre, RS, BRAZIL} 
\email{marcia.barbosa@ufrgs.br}

\date{\today}

\begin{abstract}

Using molecular dynamics simulations we investigate
the structure of a system of particles interacting  
through a continuous core-softened interparticle 
potential.  We found for the
translational order parameter, $t,$ a local 
maximum at a density $\rho_{t-max}$ and a local minimum at 
$\rho_{t-min} > \rho_{t-max}.$ Between $\rho_{t-max}$ and 
$ \rho_{t-min},$ the $t$ parameter anomalously decreases upon increasing 
pressure.
For the orientational order parameter, $Q_6$, was 
observed a maximum at a density 
$\rho_{t-max} < \rho_{Qmax} < \rho_{t-min}.$ For densities
between $\rho_{Qmax}$ and $\rho_{t-min},$ {\it both} 
the translational $(t)$ and orientational $(Q_{6})$ order
parameters have anomalous behavior.
We know that this system also exhibits density and diffusion anomaly.
We found that the region in the pressure-temperature phase-diagram 
of the structural anomaly englobes the region of the diffusion
anomaly that is larger than the region limited by the temperature
of maximum density.
 This cascade of anomalies (structural,
dynamic and thermodynamic) for our model has the same
hierarchy of that one observed for the SPC/E water.

\end{abstract}


\maketitle

%%%%%%%%%%%%%%%%%%%%%%%%%%%%%%%%%%%%%%%%%%%%%%%%%%%%%%%%%%%
\section{Introduction}
%%%%%%%%%%%%%%%%%%%%%%%%%%%%%%%%%%%%%%%%%%%%%%%%%%%%%%%%%%%

Water is the most important substance for life: 
It cools, carries, stabilizes, reacts, lubricates, dilutes, and much more.
Despite of this, many of its characteristics are not well understood.
While most liquids contract upon cooling, water expands
below $T=4^oC$ at ambient pressure \cite{Wa64}. This is known as 
the density anomaly of water.
Heating the water from $T=0^oC$ up to $T=4^oC$ 
a competition between  open low density and a closed high density
structure takes place. The gain of thermal energy 
breaks a considerable number of hydrogen bonds 
what leads  the open low density structure to
become  unstable  in relation to the closed high density structure.
So, the system contracts. 

Density anomaly is not the only one, 
far from it, the literature reports forty-one anomalies 
for water \cite{anmlies}. Not only the thermodynamics of water is 
anomalous, but 
also its dynamics. Commonly the materials diffusivity decreases 
with increasing pressure. Liquid water has an opposite behavior 
in a large region of the phase diagram 
\cite{St99,Ga96,Ha97,Sc91,Er01,Ne01,Ne02a,Ne02b,Ne02}.
Increase in pressure disturbs the structure by inclusion
of interstitial molecules that share an hydrogen bond with
another one. As a result, the bond is weakened  
and the molecule is free to move. The shared bond breaks
and the molecule by means of a small rotation, connects to
another molecule enabling the translational diffusion  \cite{Ne02a}.

Water is not an isolated case. There are other examples 
of  tetrahedrally bonded molecular liquids such as silica and
silicon \cite{An00,Sh02} that exhibit thermodynamic and dynamic anomalies.
Thermodynamic anomalies were also found in liquid metals \cite{Cu81} and 
graphite \cite{To97}. Unfortunately, a closed theory 
giving the relation between the form of the interaction potential 
and the presence of the anomalies is still missing.
 
It is reasonable to think that the structure and anomalies are deeply related.
Establishing the connection between structure and the thermodynamic
and dynamic behavior of water is a fundamental step towards understanding
the source of the anomalies.
At this point a question emerges: how can we define (measure) structure
in liquids?
Errington and Debenedetti \cite{Er01} proposed two simple metrics: 
a translational order parameter \cite{Er03}, $t,$ that  
measures the tendency of pairs of molecules to adopt 
preferential separations, and the orientational 
order parameter \cite{Er01,Ch98}, $q,$
quantifying the extend to which 
a molecule and its four nearest neighbors assume a tetrahedral arrangement.
For other crystal configurations one may use the 
orientational order  parameter introduced by Steinhardt \emph{et al.}
\cite{St83}, $Q_6,$ which depends on the number of nearest neighbors
taken into account for each molecule.
For a completely uncorrelated system (ideal gas) both $t$ and $q$ must 
to be zero and $Q_6,$  is equal to one over the square 
of the number of neighbors. For a crystal, $t,$ $q$ and  $Q_6$ are large. 
Torquato \emph{el al.} \cite{To00} introduced a systematic way to study the 
structural order in liquids mapping state points into the $t-q$ plane. They
refer to it as an order map. Errington and Debenedetti
used the order map to investigate  structural order in 
simple point charge/extended (SPC/E) water \cite{Er01}.

For normal liquids,   $t$ and $q$ increase upon compression, because
the system tends to be more structured. It was found that in SPC/E water
both $t$ and $q$ decrease upon compression in a certain region
of the pressure-temperature (P-T) phase diagram \cite{Er01}.  
This region is referred as the region of structural anomalies. 
Errington and Debenedetti  showed also that, inside 
the structurally anomalous region, all the paths formed by the $(t,q)$ 
points collapse into a single line.
This means that the translational order parameter, $t,$ and the
orientational order parameter, $q,$ are coupled.
Outside the structurally  anomalous region the states points 
in the order map define a two-dimensional
region, meaning that the parameters $t$ and $q$ are independent.

Performing molecular dynamics simulations, 
Errington and Debenedetti \cite{Er01} and 
Netz \emph{et al.} \cite{Ne01}  
showed that in SPC/E water the 
thermodynamic and dynamic anomalies form
nested domes in the P-T phase-diagram, 
where the diffusion anomaly lies outside the density anomaly.
Additionally, Errington and Debenedetti
showed that the structurally
anomalous region englobes
the diffusion and density anomalies regions.

Several models of water for computer simulations have been
proposed\cite{Gi02}, with three, four or
five localized partial charges, some of them having Lennard-Jones
interaction centers in the oxygens and hydrogens, others
only in the oxygens. A considerable number of these approaches reproduce many
anomalies present in liquid water.
However, these models are complicated, what makes difficult
to understand the physics behind the anomalies.
In this sense, isotropic models
are the simplest framework to understand the physics of
liquid state anomalies. Moreover, the use of 
an effective potential is particularly suitable for
extending our conclusions for more  complex fluids.
From the desire of constructing a simple two-body isotropic potential, 
capable of describing water-like anomalies,
a number of models in which
single component systems of particles interact via 
core-softened (CS) potentials \cite{pabloreview} have been proposed. 
They possess a repulsive core 
that exhibits a region of softening where the slope changes dramatically. 
This region can be a shoulder or a ramp 
\cite{St98,Sc00,Fr01,Bu02,Bu03,Sk04,Fr02,Ba04,Ol05,He05,He70,Ja98,Wi02,Ma04,
Ca03,Ca05,Ku04,Xu05,Zh05,Zh06,Wi06,Ol06}.

In the shoulder case, the potential consists of a hard
core, a  repulsive shoulder and, in some cases, an attractive square well
\cite{St98,Sc00,Fr01,Bu02,Bu03,Sk04,Fr02,Ba04,Ol05,He05,Ma04,Wi02,Ca03,Ca05}. 
The potential has a change in the slope at short-ranged distances.
In two dimensions, such potentials have thermodynamic and diffusion anomalies. In 
three dimensions, no dynamic and thermodynamic anomalies were reported
\cite{Fr01,Bu02,Bu03,Fr02,Sk04,Ma04}.

In the ramp  case, the interaction potential has two competing
equilibrium distances, defined by a repulsive ramp 
\cite{Ja98,Wi02,Ku04,Wi06,Zh05,Zh06,Ol06}. In some cases an attractive part
is included \cite{Ja98,Wi02,Wi06}. 
In two dimensions, there are thermodynamic anomalies in such potentials. 
In three dimensions, these potentials exhibit not only thermodynamic 
anomalies, 
but also dynamic and structural anomalies \cite{Ku04,Zh05,Zh06,Ol06}.

Notwithstanding the progresses  described above, a model
in which both the potential and the force are continuous
functions and that exhibits all the thermodynamic and dynamic anomalies 
like the ones present
in water is still missing. 
In this paper, we check if a ramp-like potential previously studied 
by us \cite{Ol06}
has not only density and diffusion anomalies, 
but also structural anomalies.
We will verify if the region in the pressure-temperature phase-diagram 
of thermodynamic and dynamic anomalies are inside the region of
structural anomalies as in SPC/E water \cite{Er01}.
The hierarchy between the anomalies in such simple model
is an important step in order to understand the mechanism
of the anomalies.

The reminder of this paper goes as follows.
In sec. II the model is introduced; in sec. III the
methods for calculating structural order in liquids
are presented. Results for the structural anomalies
and the order map obtained
from molecular dynamics simulations are shown in sec. IV.
Conclusions about the relation between the locus of 
the thermodynamic, dynamic and structural anomalies
and about the order map are presented in sec. V.

%%%%%%%%%%%%%%%%%%%%%%%%%%%%%%%%%%%%%%%%%%%%%%%%%%%%%%%%%%%
\section{The model}
%%%%%%%%%%%%%%%%%%%%%%%%%%%%%%%%%%%%%%%%%%%%%%%%%%%%%%%%%%%

The model we study consists of a system of 
$N$ particles of diameter $\sigma$  interacting through  an isotropic
effective potential given by

%%%%%%%%%%%%%%%%%%%%%%%%%%%%%%%%%%%%%%%%%%%%%%%%%%%%%%%%%%%
\begin{equation}
U^{*}(r)=4\left[\left(\frac{\sigma}{r}\right)^{12}-
\left(\frac{\sigma}{r}\right)^{6}\right]+
a\exp\left[-\frac{1}{c^{2}}\left(\frac{r-r_{0}}{\sigma}
\right)^{2}\right],
\label{eq:potential}
\end{equation}
%%%%%%%%%%%%%%%%%%%%%%%%%%%%%%%%%%%%%%%%%%%%%%%%%%%%%%%%%%

\noindent where $U^{*}(r)=U(r)/\epsilon.$
The first term of Eq. (\ref{eq:potential}) is a Lennard-Jones
potential of well depth $\epsilon$ and the second 
term is a Gaussian centered on radius $r=r_{0}$ with height $a$
and width $c$.
Depending on the choice of the values of $a$, $r_{0}$ and $c$,
this potential  assumes several shapes ranging from  
a deep double well potential \cite{Ch96,Ch97a,Ne04} 
to a repulsive shoulder \cite{Ja98}. 

\begin{figure}[ht]
\begin{center}\includegraphics[clip=true,scale=0.5]{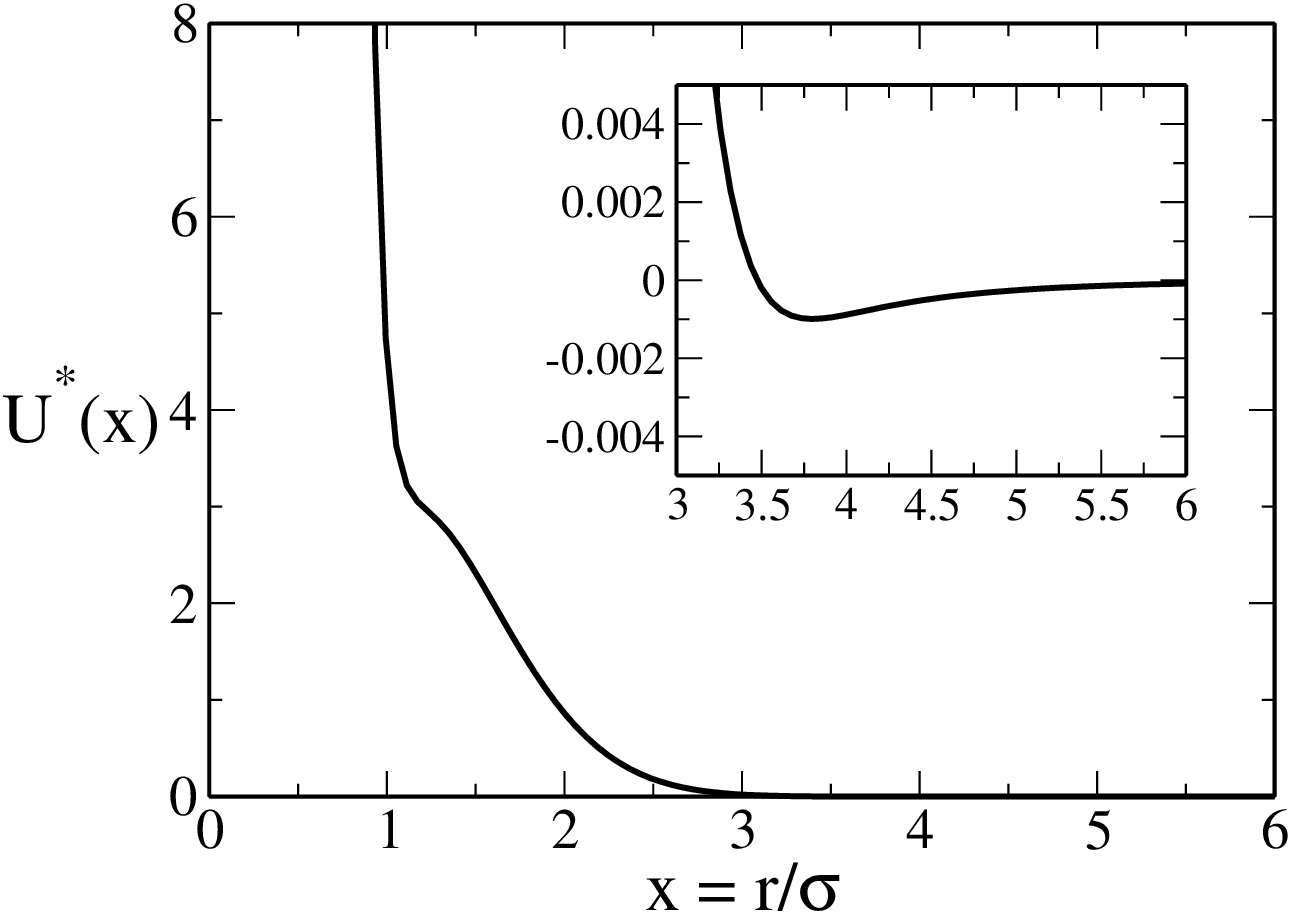}
\end{center}
\caption{Interaction potential eq. (\ref{eq:potential}) with
parameters $a=5,$ $r_{0}/\sigma=0.7$ and $c=1$, in  reduced
units. The inset shows a zoom in the very small attractive part
of the potential. \label{cap:Soft-core}}
\end{figure}

Recently, using molecular dynamics 
simulations and integral equations theory,
we have studied the potential Eq. (\ref{eq:potential}) 
setting $a=5,$ $r_{0}/\sigma=0.7$ and $c=1$ 
(see Fig. \ref{cap:Soft-core}) \cite{Ol06}. Here,
we use the same parameters (as in the previous \cite{Ol06}). It is
interesting to note that close to the core ($r/\sigma\approx1$)
this potential experiences an unusual change of slope, weakening
the repulsive force between the particles.
%%%%%%%%%%%%%%%%%%%%%%%%%%%%%%%%%%%%%%%
%%%%%%%%%%%%%%%%%%%%%%%%%%%%%%%%%%%%%%%

%%%%%%%%%%%%%%%%%%%%%%%%%%%%%%%%%%%%%%%%%%
\section{The methods}
%%%%%%%%%%%%%%%%%%%%%%%%%%%%%%%%%%%%%%%%%%

%%%%%%%%%%%%%%%%%%%%%%%%%%%%%%%%%%%%%%%%%%
\subsection{Translational order parameter}
%%%%%%%%%%%%%%%%%%%%%%%%%%%%%%%%%%%%%%%%%%

The translational order parameter of a 
system of particles of density $\rho=N/V$,
where $N$ is the number of particles and $V$ is the 
volume of the system, is defined as 
\cite{Er01,Er03,Sh02},

%%%%%%%%%%%%%%%%%%%%%%%%%%%%%%%%%%%%%%%%%
\begin{equation}
t\equiv\int_{0}^{\xi_{c}}|g(\xi)-1|d\xi,
\label{eq:trans}
\end{equation}
%%%%%%%%%%%%%%%%%%%%%%%%%%%%%%%%%%%%%%%%%

\noindent where $\xi\equiv r\rho^{1/3}$ is the interparticle distance $r$
divided by the mean separation between pairs of particles $\rho^{-1/3}. $
$g(\xi)$ is the radial distribution function, where $g$ is proportional to the
probability to find a particle at a distance $\xi$ to another particle
placed at the origin. $\xi_c$ is a cut-off distance. 
In this work, we use\cite{inspired} $\xi_c=\rho^{1/3} L/2,$
where $L=V^{1/3}.$  For a completely uncorrelated system (ideal gas) $g=1$
and $t$ vanishes. In a crystal, a translational long-order ($g\ne1$) persists
over long distances making $t$ large.

%%%%%%%%%%%%%%%%%%%%%%%%%%%%%%%%%%%%%%%%%%
\subsection{Orientational order parameter}
%%%%%%%%%%%%%%%%%%%%%%%%%%%%%%%%%%%%%%%%%%

For the orientational order parameter introduced by 
Steinhardt \emph{el. al.} \cite{St83}, we follow the 
strategy introduced by Yan \emph{el. al} \cite{Zh05}.
We define $k$ vectors, $\mathbf{r}_{ij},$ connecting the particle $i$ with
its $k$  nearest neighbors $j$. Each vector $\mathbf{r}_{ij}$
is a "bond".  A polar ($\phi_{ij}$) and azimuthal ($\theta_{ij}$) 
angles with reference to an arbitrary axis
may be associated to each bond ${r}_{ij}$ and 
the spherical harmonics $Y_{lm}(\theta_{ij},\phi_{ij})$ 
may be calculated. After computing the average of 
$Y_{lm}(\theta_{ij},\phi_{ij})$
over the $k$ bonds, namely, 

%%%%%%%%%%%%%%%%%%%%%%%%%%%%%%%%%%%%
\begin{equation}
\langle Y_{lm}^{i}\rangle =\frac{1}{k}\sum_{j=1}^{k}Y_{lm}(\theta_{ij},\phi_{ij}),
\label{eq:avYlm}
\end{equation}
%%%%%%%%%%%%%%%%%%%%%%%%%%%%%%%%%%%% 

\noindent one can evaluate the orientational order parameter 
\cite{Er01,Er03,To00,Sh02,Tr00,Hu04} associated to each particle $i$,

%%%%%%%%%%%%%%%%%%%%%%%%%%%%%%%%%%%%
\begin{equation}
Q_{l}^{i}=\left[\frac{4\pi}{2\ell+1}
\sum_{m=-\ell}^{m=\ell}\left| \langle Y_{lm}^{i}\rangle \right|^{2}\right]^{1/2}.
\label{eq:Qli}
\end{equation}
%%%%%%%%%%%%%%%%%%%%%%%%%%%%%%%%%%%%

For characterize the local order \cite{orders} of the system 
was used \cite{Zh05,Zh06} 

%%%%%%%%%%%%%%%%%%%%%%%%%%%%%%%%%%%%
\begin{equation}
Q_6=\frac{1}{N}\sum_{i=1}^{N}Q_{6}^{i},
\label{eq:Q6}
\end{equation}
%%%%%%%%%%%%%%%%%%%%%%%%%%%%%%%%%%%%

\noindent that is the mean value of $Q_{6}^{i}$ over all particles 
of the system. The $Q_6$ parameter assumes its maximum value for a 
perfect crystal and decreases as the system becomes less structured.
For a completely
uncorrelated system (ideal gas) $Q_{6}^{ig}=1/\sqrt{k}.$
For a crystal, the $Q_6$ value depends on
the specific crystalline arrangement and 
the number of neighbors taken into account. 
For example, for the face centered cubic (fcc) with
its twelve first neighbors ($k=12$), 
we have $Q_{6}^{fcc}=0.574.$ For a body
centered cubic (bcc), which have only eight
nearest neighbors, $Q_{6}^{bcc-8}=0.628.$
Note that if we include not eight, but fourteen
neighbors for calculating $Q_{6}^{bcc-k}$, we have 
$Q_{6}^{bcc-14}=0.510.$

For the potential given by the Eq. (\ref{eq:potential}), the
expected crystalline configuration at the ground state 
for low densities is 
the hexagonal close packing (hcp), which have
twelve first neighbors (see sec. \ref{secresults} for more details).
In this work we used $k=12$ in the Eq. (\ref{eq:avYlm}). For
the hcp crystal, $Q_{6}^{hcp}=0.484$.

%%%%%%%%%%%%%%%%%%%%%%%%%%%%%%%%%%
\section{Results from simulations \label{secresults}}
%%%%%%%%%%%%%%%%%%%%%%%%%%%%%%%%%%

We performed molecular dynamics (MD)  simulations in the
canonical ensemble using 500 particles in a cubic box
with periodic boundary conditions, interacting with 
the potential Eq. (\ref{eq:potential}). The parameters
employed were $a=5,$ $r_0/\sigma=0.7,$ and $c=1.0.$ The cutoff
radius was set to $3.5 \sigma$\cite{paper}. In order to keep fixed
the temperature, the Nos\'e-Hoover
\cite{Nh85} thermostat was used with the coupling constant $q_{NH}=2.$ 
Pressure, temperature, and density are shown in dimensionless units,

%%%%%%%%%%%%%%%%%%%%%%%%%
\begin{equation}
P^{*} \equiv \frac{P \sigma^3}{\epsilon}
\label{eq:PP}
\end{equation}
%%%%%%%%%%%%%%%%%%%%%%%%%

%%%%%%%%%%%%%%%%%%%%%%%%%
\begin{equation}
T^{*} \equiv \frac{k_{B} T}{\epsilon}
\label{eq:TT}
\end{equation}
%%%%%%%%%%%%%%%%%%%%%%%%%

%%%%%%%%%%%%%%%%%%%%%%%%%
\begin{equation}
\rho^{*} \equiv \rho \sigma^3.
\label{eq:RR}
\end{equation}
%%%%%%%%%%%%%%%%%%%%%%%%%

The translational and orientational order
parameters were calculated over 1 000 000 steps MD simulations, 
previously equilibrated over 200 000 steps. For low temperatures
($T^{*}<0.4$), additional simulations were carried out
with equilibration over 500 000 steps, followed by 
2 000 000 steps simulation run.
The time step was 0.002 in reduced units.  

\begin{figure}[ht]
\begin{center}\includegraphics[clip=true,scale=0.6]{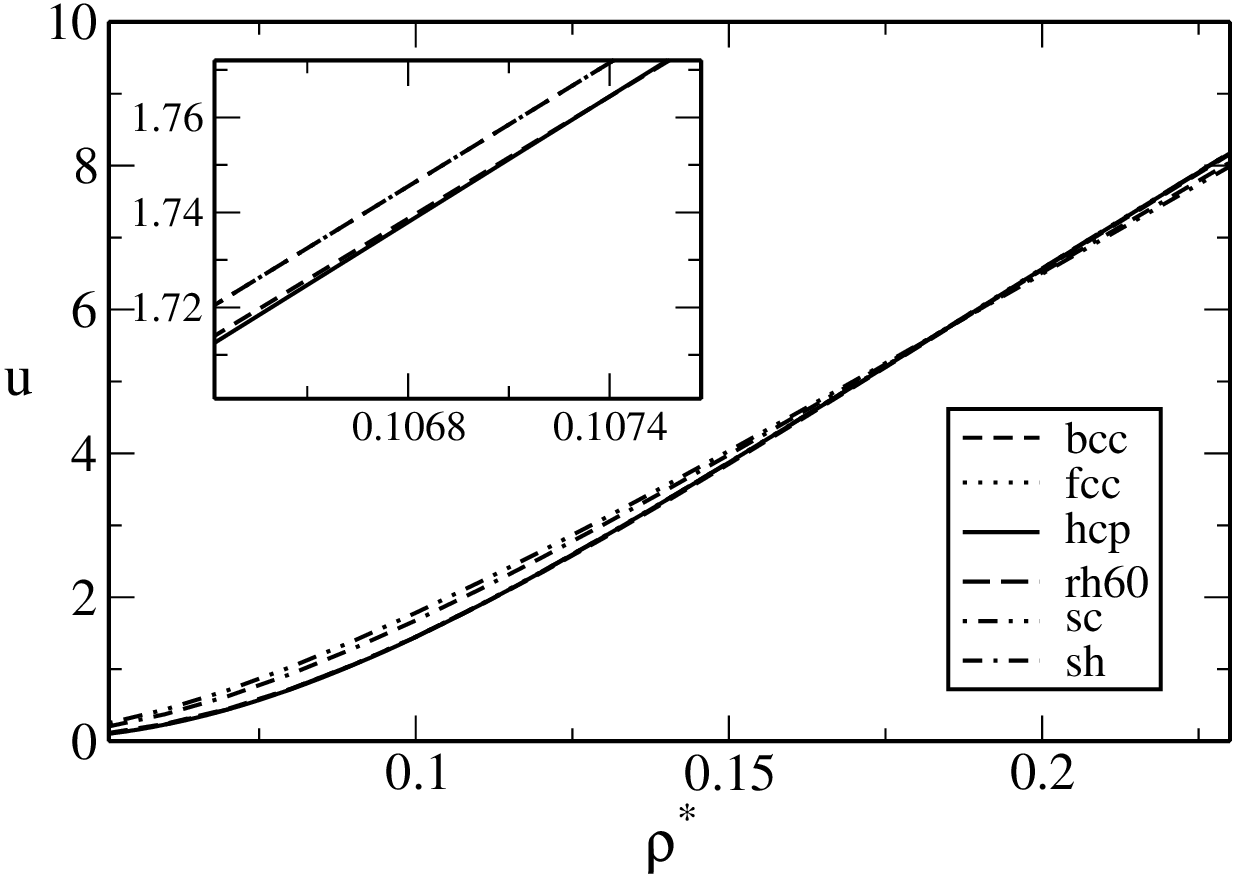}
\end{center}
\caption{ The dimensionless configurational energy per particle for 
the several crystal structures considered: face centered cubic (fcc), 
body centered cubic (bcc), simple cubic (sc), simple hexagonal (sh),
hexagonal closest packing (hcp), and rhombohedral-60$^o$ (rh60).
We see that the hcp has the lower configuration energy per particle
for densities $\rho^{*}\lesssim0.107$ (see inset). 
Hence, the expected structure for our model at
$T^{*}=0$ is the hcp for $\rho^{*}\lesssim0.107.$ 
For $0.107\lesssim\rho^{*}\lesssim0.187$ the bcc
phase has the lower configurational energy between those studied
(not shown).
\label{cap:energies}}
\end{figure}

For studying the crystalline structure 
of our model we consider
the expected, following conformations for the ramp potential \cite{Ja98}: 
simple cubic (sc), body centered cubic (bcc), face centered cubic 
(fcc), simple hexagonal (sh),  hexagonal close packing (hcp), 
and the rhombohedral-$60^o$ (rh60). Perfect crystals with such 
conformations were constructed and the configurational
energy per particle $u=U^{*}/N$ was
calculated for each arrangement.
In the canonical ensemble, the 
most stable crystal at the ground state is that one with
lower $u$. From the Fig. \ref{cap:energies}, we see that the hcp
conformation is the more stable for densities
$\rho^{*}\lesssim0.107$ (see inset). The bcc 
conformation is the more stable one for  
$0.107\lesssim\rho^{*}\lesssim0.187$ (not shown).

%%%%%%%%%%%%%%%%%%%%%%%%%%%%%%%%%%%%%%%
%%%%%%%%%%%%%%%%%%%%%%%%%%%%%%%%%%%%%%%

%%%%%%%%%%%%%%%%%%%%%%%%%%%%%%%%%%%%%%%
\begin{figure}[ht]
\begin{center}\includegraphics[clip=true,scale=0.6]{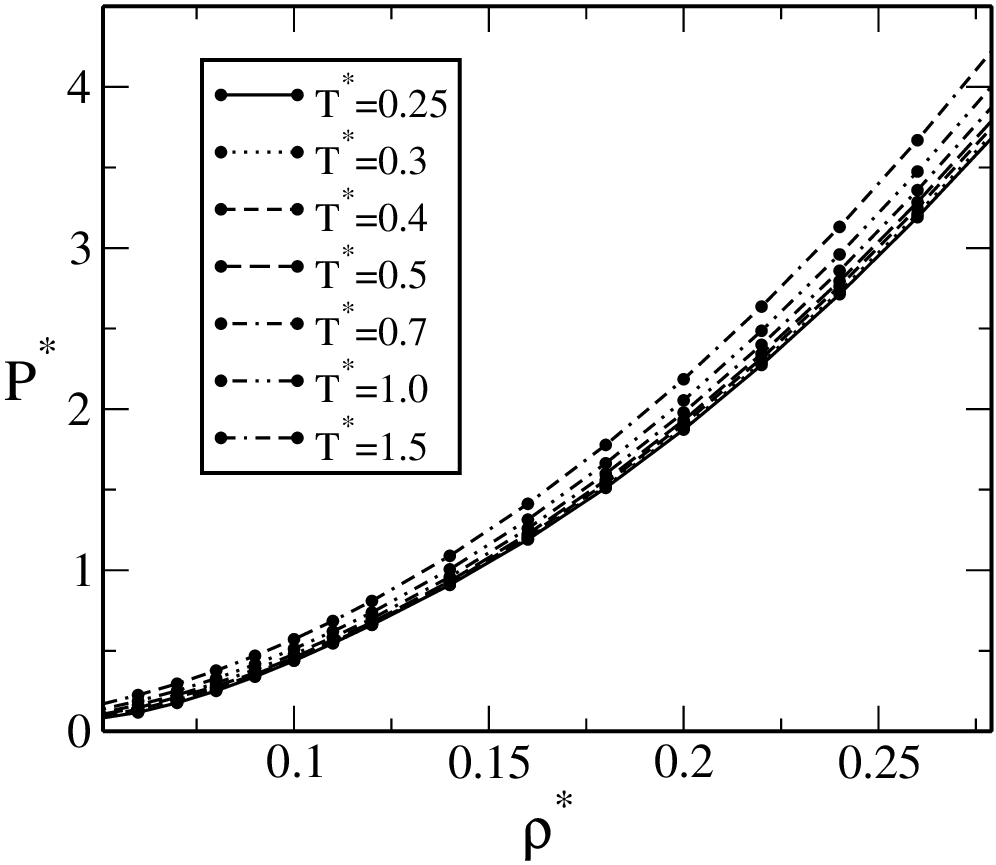}
\end{center}
\caption{The reduced pressure as a function of the reduced density.
The seven isotherms show that the relation between $P^*$ and
$\rho^*$ is monotonic.
\label{cap:figprho}}
\end{figure}
%%%%%%%%%%%%%%%%%%%%%%%%%%%%%%%%%%%%%%%

Studying the equation of state pressure against density for our model 
we found a monotonic behaviour for $P(\rho).$ Hence,
an increase in pressure means increase in density as shown in
Fig. \ref{cap:figprho}.

\begin{figure}[ht]
\begin{center}\includegraphics[clip=true,scale=0.6]{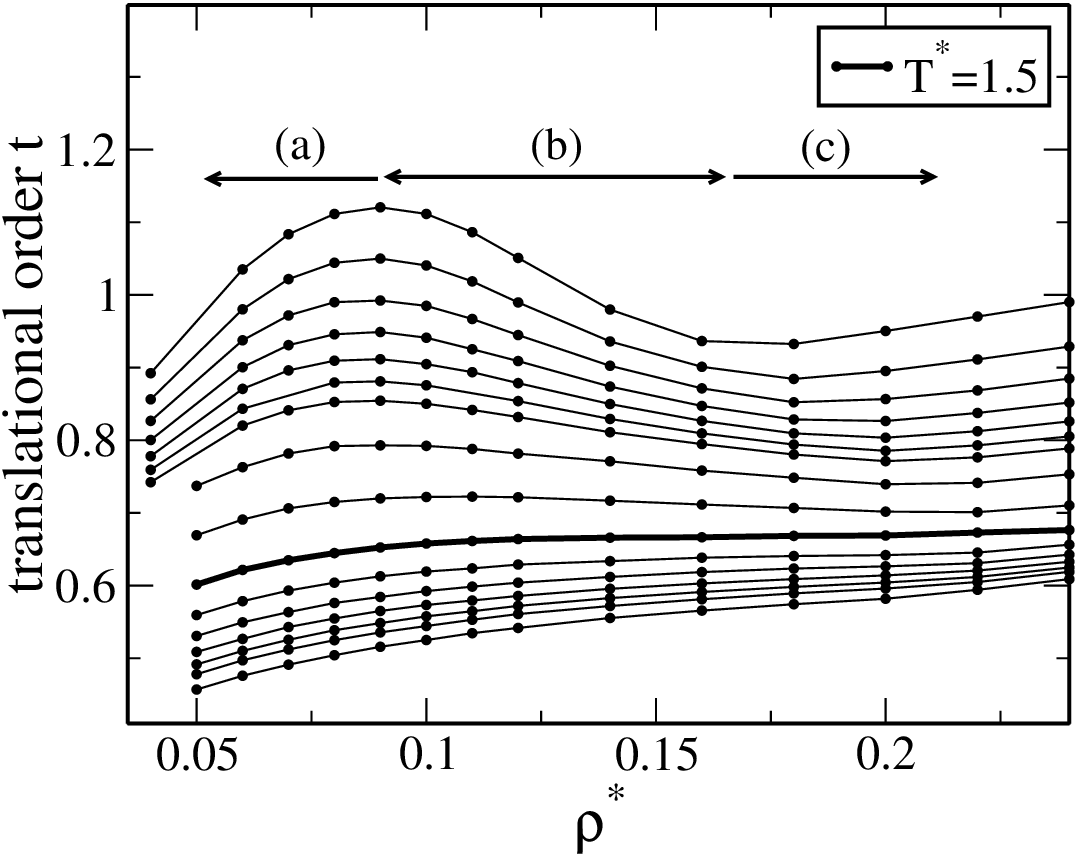}
\end{center}
\caption{The translational order parameter $t$ as a function of
the density $\rho^{*}.$  From top to bottom, the sixteen isotherms
are $T^{*}=$ 0.25, 0.30, 0.35, 0.40, 0.45, 0.50, 0.55, 0.70, 1.0,
1.5, 2.0, 2.5, 3.0, 3.5, 4.0, and 5.0. 
The (a), (b), and (c) arrows corresponds to the density
range spanned by the Fig. \ref{cap:grs}(a), \ref{cap:grs}(b), and
\ref{cap:grs}(c) respectively.
The bold line indicates
the isotherm $T^{*}=1.5.$ For $T^{*}>1.5$ no anomalous behavior
is observed for $t.$ The line connecting the points are just a guide
for the eyes.
\label{cap:figt}}
\end{figure}

\begin{figure}[ht]
\center
\includegraphics[clip=true,scale=0.7]{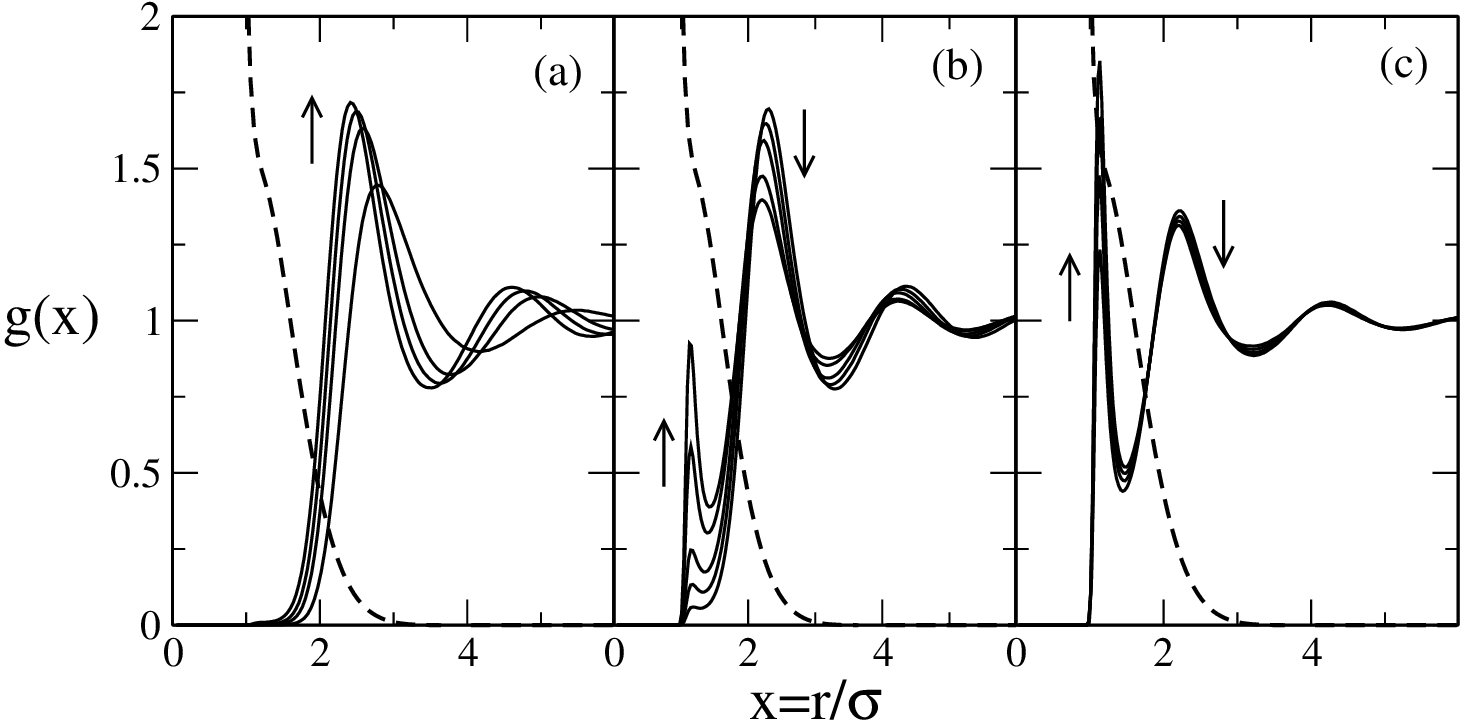}
\caption{The radial distribution functions for $T^{*}=0.25$ 
and several densities: (a) $\rho^{*}=$ 0.04, 0.06, 0.07, and 0.08;
(b) $\rho^{*}=$ 0.10, 0.11, 0.12, 0.14, and 0.16; (c) $\rho^{*}=$
0.18, 0.20, 0.22, and 0.24. The arrows indicate the direction of
increasing $\rho^{*}.$ The dashed
line is the reduced interparticle potential shown in
the Fig. \ref{cap:Soft-core}
multiplied by a factor of 0.5 just for clarity.
\label{cap:grs}}
\end{figure}

Results for the translational  order parameter
for the liquid phase can be seen
in the Fig. \ref{cap:figt}. While for
a normal liquid $t$ increases under compression,
for our system this is the case only for
high temperatures. For lower temperatures $t$ presents a local maximum 
at a density $\rho_{t-max}$ and a local minimum at a density 
$\rho_{t-min} > \rho_{t-max}$ 
for temperatures $T^{*}<1.5.$
Between $\rho_{t-max}$ and $\rho_{t-min}$  an unusual behavior for 
the translational order 
parameter is observed: An increase in density induces a decrease in 
translational order. 
This behavior can be understood analyzing the dependence of  the
radial distribution function (RDF) upon density [see Eq. (\ref{eq:trans})].
The (a), (b), and (c) arrows in the Fig. \ref{cap:figt} 
correspond to the density range 
spanned by the Fig. \ref{cap:grs}(a), \ref{cap:grs}(b), and
\ref{cap:grs}(c) respectively.

The Fig. \ref{cap:grs}  shows the RDF for $T^{*}=0.25$ and several densities: 
(a) $\rho^{*}=$ 0.04, 0.06, 0.07 and 0.08; (b) 
$\rho^{*}=$ 0.1, 0.11, 0.12, 0.14, 
and 0.16; (c) $\rho^{*}=$ 0.18, 0.2, 0.22, and 0.24. The arrows indicate
the directions of increasing $\rho^{*}$ and the dashed
line is the reduced interparticle potential shown in
the Fig. \ref{cap:Soft-core} 
multiplied by a factor of 0.5 just for clarity. 
From Fig. \ref{cap:grs}(a) we
see the growth
of $g(r)$ at $r/\sigma\approx 2.5$ upon compression,
causing an increasing of $t$ over the range 
$0.04\leq\rho^{*}\leq0.08.$ See the isotherm $T^{*}=0.25$
at Fig. \ref{cap:figt}.  
In this range of densities the particles are repelled
by the repulsive shoulder and the most probable separation is
about 2.5 units (corresponding to the edge of the shoulder
in the potential).  
We see that over the intermediate density range
$0.08<\rho^{*}<0.18,$ $t$ decreases as the density increases. 
Looking at the Fig. \ref{cap:grs}(b) one can explain why this happens.
Both an increase of $g(r)$ at $r/\sigma\approx1.0,$ approximating 
the RDF to 1 (what decreases $|g(r)-1|$), and a decrease of $g(r)$
in the next peak (close to 2.5) upon compression, causes $t$  to decrease.
This new peak at about 1 unit corresponds to the position of the
hard-core part of the potential \cite{clustering}.
Finally, $t$ returns to increase upon compression for $\rho^{*}>0.18.$ 
The sharp  growth of $g(r)$ at $r/\sigma\approx1.0$ 
above the unity [see Fig.\ref{cap:grs}(c)] 
underlies this behavior, indicating that all the particles
are pushed together up to their hard-cores.
This was the same behavior observed for the RDF of the ramp potential \cite{Zh05,Zh06}.
Anomalous variations in $t$ are absent for $T^{*}>1.5$ because the thermal energy washes
out the effect of the repulsive shoulder.

%%%%%%%%%%%%%%%%%%%%%%%%%%%%%%%%%%%%%%%
%%%%%%%%%%%%%%%%%%%%%%%%%%%%%%%%%%%%%%%
%%%%%%%%%%%%%%%%%%%%%%%%%%%%%%%%%%%%%%%%%%%%%
%%%%%%%%%%%%%%%%%%%%%%%%%%%%%%%%%%%%%%%%%%%%%

\begin{figure}[ht]
\begin{center}\includegraphics[clip=true,scale=0.6]{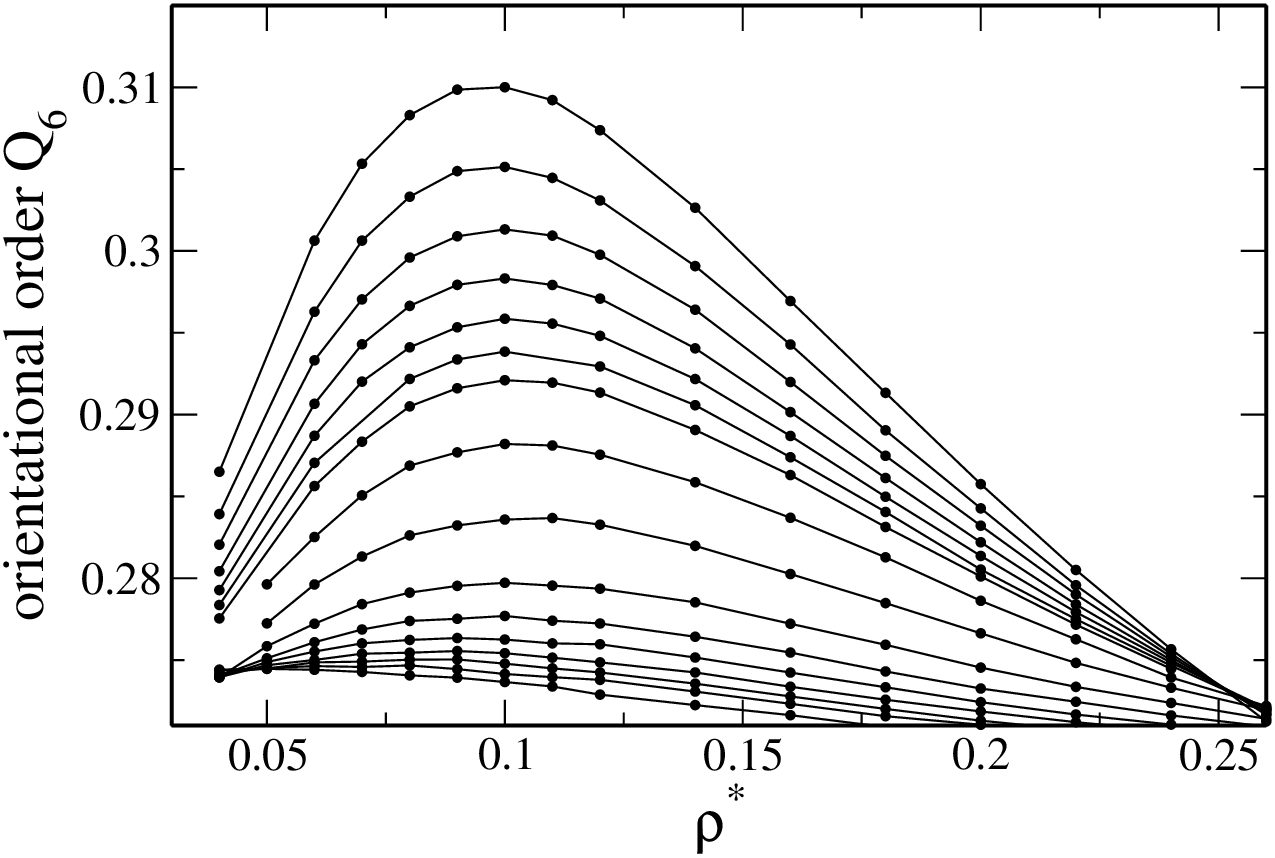}
\end{center}
\caption{The orientational order parameter $Q_6$ as a function of
the density $\rho^{*}.$  From top to bottom, the sixteen isotherms
are $T^{*}=$ 0.25, 0.30, 0.35, 0.40, 0.45, 0.50, 0.55, 0.70, 1.0,
1.5, 2.0, 2.5, 3.0, 3.5, 4.0, and 5.0. For $5.0<T^{*}<8.0$ (not shown)
the $Q_6$ local maximum points  occur at $\rho^{*}=0$ with
a global minimum at $\rho^{*}\approx 0.3.$ 
We do not study the cases where $T^{*}>8.0$ (see the text for more details).
The line connecting the points are just a guide for the eyes.
\label{cap:figq}}
\end{figure}

For a normal liquid, it is expected that the
orientational  order parameter,  $Q_6$, increases under 
compression. For our potential, however, 
a local maximum is detected for $Q_6$ at a density 
$\rho_{Qmax}$ in such a way that $\rho_{t-max}<\rho_{Qmax}<\rho_{t-min}$ 
(see figure   \ref{cap:figq}).
This means that for densities between $\rho_{Qmax}$ and $\rho_{t-min}$
{\it both} the structural order parameters
$t$ and $Q_6$ have an anomalous behavior, since
$t$ and $Q_6$ decrease under increasing of  pressure. 
We call this range of densities  the structural anomaly domain.

%%%%%%%%%%%%%%%%%%%%%%%%%%%%%%%%%%%%%%%
%%%%%%%%%%%%%%%%%%%%%%%%%%%%%%%%%%%%%%%

\begin{figure}[ht]
\center
\includegraphics[clip=true,scale=0.6]{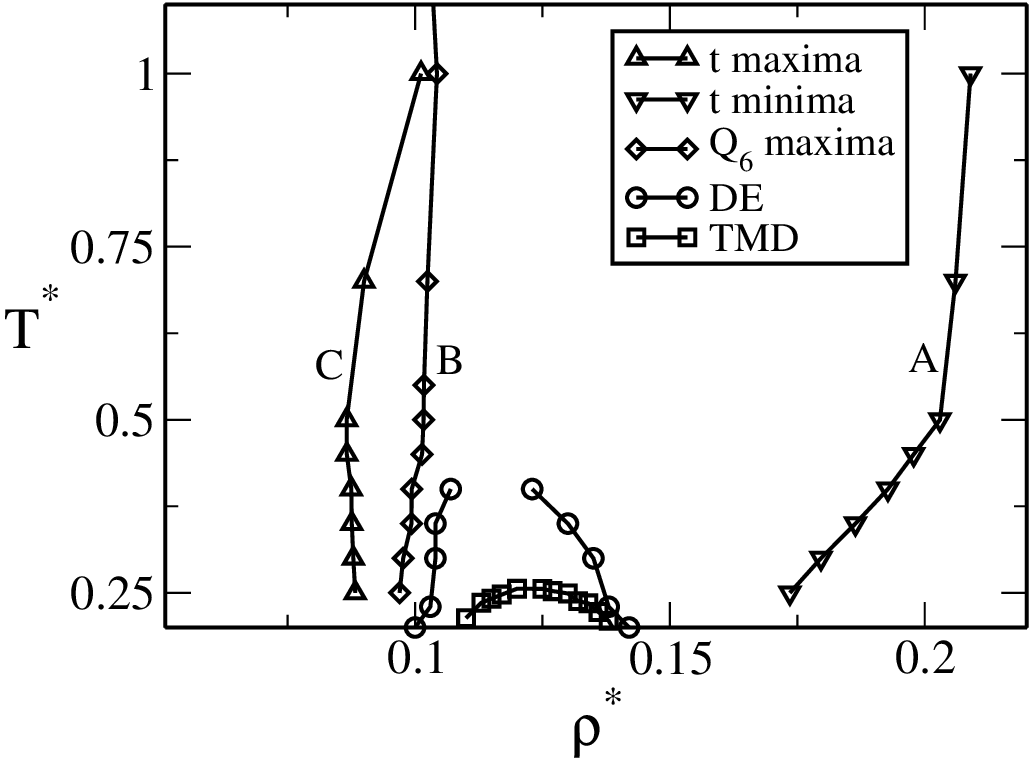}
\caption{The relationship between the several anomalies presented
for our model. The curve B ends at $T^{*}\approx5.0$
and it is not entirely shown for clarity. See the text for more details.
Between the $Q_6$ maxima line (curve B) and t minima line
(curve A) {\it both} the translational and orientational order 
parameters $t$ and 
$Q_6$ become anomalous, namely, decrease with density. This region we call
structural anomaly region. The diffusion extrema (DE) lines enclose the region
inside which the diffusion decreases with density $-$ the dynamic 
anomaly region.
The temperature of maximum density (TMD) line englobes the region that 
density anomaly appears.
Both the DE and TMD lines were obtained from previous work \cite{Ol06}. These 
cascade 
of anomalies presents the same hierarchy as observed for the SPC/E 
water \cite{Er01,Ne01}.
\label{cap:cascade}}
\end{figure}

The relation between the several anomalies presented for this potential
is shown in the Fig. \ref{cap:cascade}.  The temperature of maximum
densities (TMD) and
the diffusivity extrema (DE) lines were obtained from previous 
work \cite{Ol06}.
The TMD line indicates the region of thermodynamic anomaly region, inside
which the density increases when the system is heated at constant pressure.
The DE lines determinate the region of dynamic anomaly.
Inside this region, diffusivity increases with increasing density.
In this work we determinate additional
three lines shown at Fig. \ref{cap:cascade}: the curve of
$t$ maxima  (C), the curve of   $Q_6$ maxima  (B), and the 
curve of  $t$ minima (A). We call
the region between the  curves A and B the structural anomalous region,
inside which both the order parameters, $t$ and $Q_6,$ become anomalous,
namely, decrease with density. The curve B, composed by the $Q_6$ local maxima
points, terminates at $T^{*}\approx5.0,$ not shown in the 
Fig. \ref{cap:cascade}
for clarity. As the temperature $T^{*}$ tends to 5.0, the densities
for the $Q_6$ maxima loci tends to zero. For $T^{*}>5.0$ we have studied
the temperatures $T^{*}=$ 5.5, 6.0, 6.5, 7.0, and 8.0 (not shown). 
For all these temperatures
the same qualitative  behavior for $Q_6$ was observed: The $Q_6$ parameter
has local maxima at $\rho^{*}=0$ and global minima at $\rho^{*}\approx0.3.$
The ratio between these extrema (local maxima and global minima) 
does not extrapolate $3.5\%$ in any case.
We do not simulate temperatures $T^{*}>8.0.$

%%%%%%%%%%%%%%%%%%%%%%%%%%%%%%%%%
%%%%%%%%%%%%%%%%%%%%%%%%%%%%%%%%%

%%%%%%%%%%%%%%%%%%%%%%%%%%%%%%%%%
\begin{figure}[ht]
\center
\includegraphics[clip=true,scale=0.6]{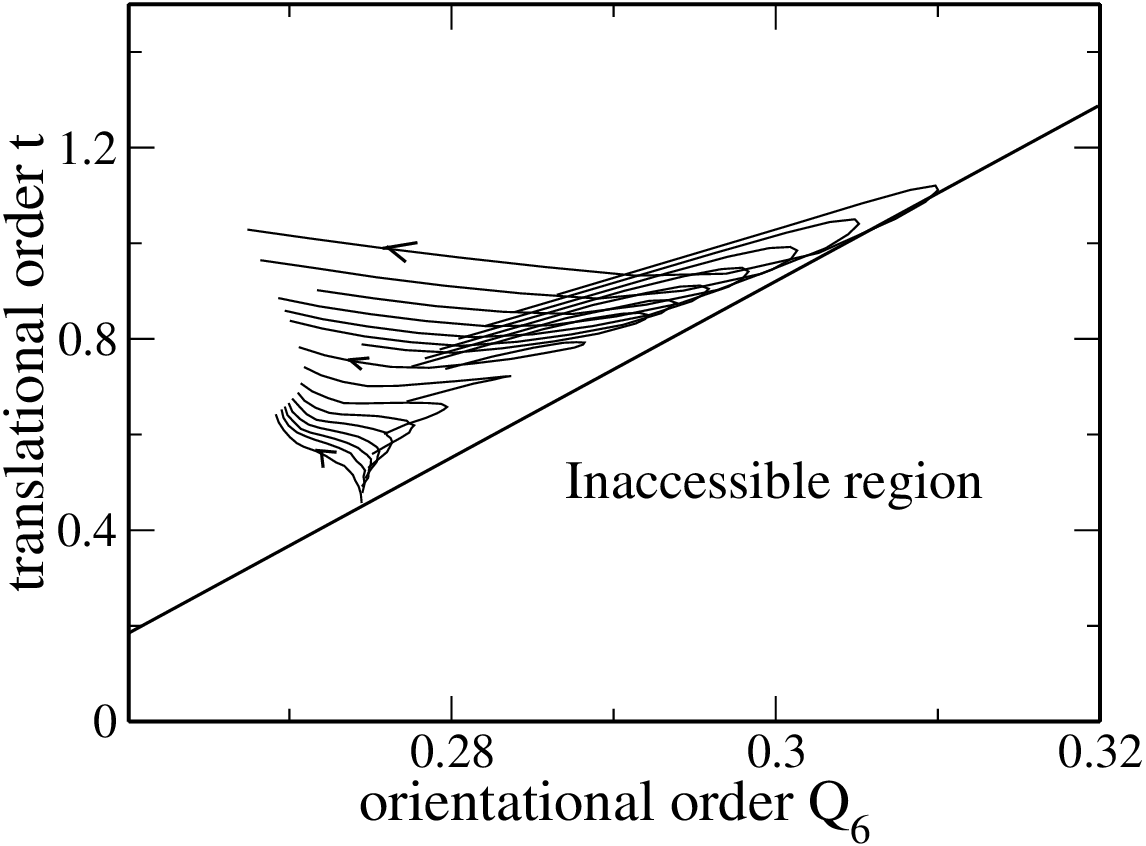}
\caption{The $t-Q_{6}$ plane, or order map. 
Each line corresponds to an isotherm and
the arrows indicate the direction of density growth.
From top to bottom,
the isotherms showed here are: $T^{*}=$ 0.25, 0.3, 0.35, 0.4,
0.45, 0.5, 0.55, 0.7, 1.0, 1.5, 2.0, 2.5, 3.0, 3.5, 4.0, and
5.0. Unlike the SPC/E water \cite{Er01},
the paths formed by the $t$ and $Q_6$ parameters developed a two dimensional
region in the order map for temperatures and densities inside the structural
anomalous region. As observed for the SPC/E water \cite{Er01}.
\label{cap:ordermap}}
\end{figure}
%%%%%%%%%%%%%%%%%%%%%%%%%%%%%%%%%

For the SPC/E water \cite{Er01,Ne01}, the region of structural anomalies
contains (inside) the region of dynamic anomalies, and the thermodynamic anomaly 
region lies inside this last one.  For the silica, other
tetrahedrally bonded molecular liquid, simulations show \cite{Sh02}
an inverse order between the structural and dynamic anomaly regions: The 
diffusion
anomaly region englobes the structural anomaly region that
englobes the thermodynamic anomaly region. For our model,
we see a water-like cascade of anomaly regions similar to that found for the 
SPC/E water (see Fig. \ref{cap:cascade}). This suggests that the 
role played by the structure in our potential, like in water,
is determinant for giving rise to the other anomalies.

From the Fig. \ref{cap:energies} its clear that for densities over
$0.107$ the hcp conformation is not the most stable one
between those studied. The bcc configuration becomes the expected
crystalline arrangement for densities 
$0.107\lesssim\rho^{*}\lesssim0.187.$
Hence, the calculation of the $Q_6$ parameter was repeated
using eight first neighbors ($k=8$) in the Eq. (\ref{eq:avYlm}).
For this new calculation, the entire curve B in the Fig. \ref {cap:cascade}
is shifted approximately $13\%$ (not shown) in the direction of lower densities,
crossing the curve C. In this new scenario, the structural anomaly region now
lies between the curves C and A, region 
in which both $Q_6$ and $t$ decrease upon compression. 
Despite of this change, the whole
qualitatively result is not modified, once the structural anomaly region
remains outside the dynamic and thermodynamic regions.

As discussed in the introduction,
the convenient orientational order parameter for
tetrahedral liquids \cite{Er01,Sh02} is $q.$
It was reported that for SPC/E water
the isothermal paths in a $t-q$ diagram ordermap collapse
into a single line in the structural anomaly region \cite{Er01}. 
This property supports the idea 
that in water the anomalies in  translational diffusion and in
rotational mobility are related \cite{Ne02a,Ne02b,Ne02}.

In order to check if $t$ and $Q_6$ are also related in our isotropic
model, the  order map  was also constructed. The
Fig. \ref{cap:ordermap}
shows the behavior of $t$ as a function of $Q_{6}.$ The arrows
indicate the growth of density for each isotherm.
Similar to the results found for the SPC/E water \cite{Er01}, 
Silica \cite{Sh02}, and for the ramp potential \cite{Zh05,Zh06},
was observed an inaccessible region for the order map of our model. 
However, differently from the SPC/E water \cite{Er01}, and similarly to the 
ramp potential \cite{Zh05,Zh06},
the parameters $t$ and $Q_{6}$ do not fall into a straight line  
in the order map for densities and temperatures 
inside the region of structural anomalies (note in the Fig. \ref{cap:ordermap}
that $ t$ and $Q_6$ develop a two dimensional region in the order map).

%%%%%%%%%%%%%%%%%%%%%%%%%%%%%%%%%%%%%%%%%%
\section{Conclusions}
%%%%%%%%%%%%%%%%%%%%%%%%%%%%%%%%%%%%%%%%%%

Using molecular dynamic simulations we have studied the structure of fluids
interacting via a three-dimensional continuous core-softened
potential with a continuous force. The 
translational ($t$) and orientational 
($Q_6$) order parameters introduced by
Steinhardt \emph{el. al.} \cite{St83} were analyzed in the 
framework proposed by
Yan \emph{el. al} \cite{Zh05}
to quantify the structure order for an isotropic liquid.

Our  model
exhibits   a region of density anomaly, inside which
the density increases as the system is heated at constant pressure,
and a region of diffusion anomaly, where the diffusivity
decreases with increasing  density \cite{Ol06}. In the 
pressure-temperature phase diagram, the density 
anomaly region lies inside the diffusion anomaly one.

Complementary to the thermodynamic and dynamic
anomalies,  both $t$ and $Q_6$ behave anomalously
in a large region of the temperature$-$density plane, as 
follows. The parameter $t$ have both a local maximum, at a density 
$\rho_{t-max},$ and a local minimum at a density $\rho_{t-min}>\rho_{t-max}.$
For densities in the range $\rho_{t-max}<\rho<\rho_{t-min}$ 
the translational order parameter decreases under pressure.
For normal liquids the opposite behavior is expected.
For the parameter $Q_6,$ a maximum at a density $\rho_{Qmax}$ between
$\rho_{t-max}$ and $\rho_{t-min}$ was observed. 
Hence, {\it both} $t$ and $Q_6$ become anomalous
for densities in the range $\rho_{Qmax}<\rho<\rho_{t-min}.$
The loci of the  $Q_6$ maxima, $t$ maxima, and $t$ minima
were plotted in a temperature$-$density plane and 
we showed that the region where $t$ and $Q_6$
behave anomalously encloses the regions
of density and diffusion anomalies discussed above.
This is the same behavior observed for the SPC/E water
\cite{Er01,Ne01}.
Differently from SPC/E water,
the parameters $t$ and $Q_{6}$ do not fall into a straight line  
in the order map for densities and temperatures 
what suggests that unlike water $t$ and $Q_{6}$ are independent
in the anomalous region.

In resume, the studied continuous core-softened pair potential, despite
not having long-ranged or directional interactions, exhibits
thermodynamic, dynamic \cite{Ol06}, and structural anomalies
similar to the ones observed in SPC/E water \cite{Er01,Ne01}. Therefore, we
can conclude that the presence of  anisotropy in
the interaction potential  is not a requirement for
the presence of thermodynamic, dynamic and structural 
anomalies.

%%%%%%%%%%%%%%%%%%%%%%%%%%%%%%
\subsection*{Acknowledgments}
%%%%%%%%%%%%%%%%%%%%%%%%%%%%

We thank the Brazilian science agencies CNPq, FINEP  and Fapergs 
for financial support.

\newpage


\begin{thebibliography}{10}


\bibitem{Wa64} R. Waller, Essays of Natural Experiments, Johnson Reprint corporation, New York , 1964.

\bibitem{anmlies}$<$http://www.lsbu.ac.uk/water/anmlies.html$>$.

\bibitem{Ne02a} P. A. Netz, F. W. Starr, M. C. Barbosa and
H. E. Stanley, Physica A {\bf 314}, 470 (2002).

\bibitem{St99}F. W. Starr, F. Sciortino and H. E. Stanley, Phys. Rev. E
{\bf 60}, 6757 (1999); F. W. Starr, S. T. Harrington, F. Sciortino and
H. E. Stanley, Phys. Rev. Lett. {\bf 82}, 3629 (1999).

\bibitem{Ga96}P. Gallo, F. Sciortino, P. Tartaglia and
S. -H. Chen, Phys. Rev. Lett. {\bf 76}, 2730 (1996);
F. Sciortino, P. Gallo, P. Tartaglia and S. -H. Chen,
Phys. Rev. E {\bf 54}, 6331 (1996);
S. -H. Chen, P. Gallo, F. Sciortino and P. Tartaglia, \emph{ibid.}
{\bf 56}, 4231 (1997); F. Sciortino, L. Fabbian, S. -H. Chen
and P. Tartaglia \emph{ibid.} {\bf 56}, 5397 (1997).

\bibitem{Ha97}S. Harrington, P. H. Poole, F. Sciortino and
H. E. Stanley, J. Chem. Phys. {\bf 107}, 7443 (1997).

\bibitem{Sc91}F. Sciortino, A. Geiger and H. E. Stanley, Nature (London)
{\bf 354}, 218 (1991); J. Chem Phys. {\bf 96}, 3857 (1992).

\bibitem{Er01}J. R. Errington and P. G. Debenedetti, Nature (London)
{\bf 409}, 318 (2001). 

\bibitem{Ne01}P. A. Netz, F. W. Starr, H. E.  Stanley and
M. C. Barbosa, J. Chem. Phys. {\bf 115}, 344 (2001).


\bibitem{Ne02b} H. E. Stanley, M. C. Barbosa, S. Mossa, P. A. Netz,
F. Sciortino, F. W. Starr and M. Yamada, Physica A {\bf 315}, 281
(2002).

\bibitem{Ne02} P. A. Netz, F.W. Starr, M. C. Barbosa and 
H. E. Stanley, J. Mol. Liquids {\bf 101}, 159 (2002).



\bibitem{An00} C. A. Angell, R. D. Bressel, M. Hemmatti, E. J. Sare, and
J. C. Tucker, Phys. Chem. Chem. Phys. {\bf 2}, 1559 (2000).

\bibitem{Sh02}M. S. Shell, P. G. Debenedetti, and A. Z. Panagiotopoulos, Phys. Rev. E {\bf 66}, 011202 (2002).


\bibitem{Cu81}P. T. Cummings and G. Stell, Mol. Phys. {\bf 43}, 1267 (1981).

\bibitem{To97}M. Togaya, Phys. Rev. Lett. {\bf 79}, 2474 (1997).

\bibitem{Er03}J. R. Errington, P. G. Debenedetti, and S. Torquato, J. Chem. Phys.
{\bf 118}, 2256 (2003).

\bibitem{Ch98}P. L. Chau and A. J. Hardwick, Mol. Phys. {\bf 93}, 511 (1998).

\bibitem{St83}P. J. Steinhardt, D. R. Nelson, and M. Ronchetti, Phys. Rev. B {\bf 28}, 784 (1983).



\bibitem{To00}S. Torquato, T. M. Truskett, and P. G. Debenedetti, Phys. Rev. Lett. 
{\bf 84}, 2064 (2000).

\bibitem{Gi02}B. Guillot, J. Mol. Liquids {\bf 101}, 219 (2002).

\bibitem{pabloreview} For a recent review, see 
P. Debenedetti, J. Phys.: Condens. Matter {\bf 15}, R1669 (2003).

\bibitem{St98} M. R. Sadr-Lahijany, A. Scala, S. V. Buldyrev, and
H. E. Stanley, Phys. Rev. Lett. {\bf 81}, 4895 (1998);
M. R. Sadr-Lahijany, A. Scala, S. V. Buldyrev, and H. E. Stanley,
Phys. Rev. E {\bf 60}, 6714 (1999).

\bibitem{Sc00}A. Scala, M. R. Sadr-Lahijany, N. Giovambattista,
S. V. Buldyrev and H. E. Stanley, J. Stat. Phys. 
{\bf 100}, 97 (2000); A. Scala, M. R. Sadr-Lahijany, N. 
Giovambattista, S. V. Buldyrev
and H. E. Stanley, Phys. Rev. E {\bf 63}, 041202 (2001).

\bibitem{Fr01} G. Franzese, G. Malescio, A. Skibinsky, S. 
V. Buldyrev and H. E. Stanley, Nature {\bf 409}, 692 (2001).


\bibitem{Bu02} S. V. Buldyrev, G. Franzese, N. Giovambattista,
G. Malescio, M. R. Sadr-Lahijany, A. Scala, A. Skibinsky, H. E. 
Stanley, Physica A {\bf 304}, 23 (2002).

\bibitem{Bu03} S. V. Buldyrev and H. E. Stanley,
Physica A {\bf 330}, 124 (2003).

\bibitem{Fr02} G. Franzese, G. Malescio, A. Skibinsky, S. 
V. Buldyrev and H. E. Stanley, Phys. Rev. E {\bf 66}, 051206 
(2002).

\bibitem{Ba04} Aline Balladares and Marcia C. Barbosa, J. Phys.: Cond. Matt. 
{\bf 16}, 8811 (2004).

\bibitem{Ol05} Alan B. de Oliveira and Marcia C. Barbosa, J. Phys.: Cond. Matt.
{\bf 17}, 399 (2005).

\bibitem{He05} V. B. Henriques and M. C. Barbosa, Phys.  
Rev. E {\bf 71}, 031504 (2005); V. B. Henriques, Nara Guissoni, Marco 
A. Barbosa, Marcelo Thielo and Marcia C. Barbosa, Mol. Phys. {\bf 103},
3001 (2005).

\bibitem{Sk04} A. Skibinsky, S. V. Buldyrev, G. Franzese, G. Malescio 
and H. E. Stanley, Phys. Rev. E {\bf 69}, 061206 (2004).


\bibitem{Ma04}G. Malescio, G. Franzese, A. Skibinsky, S. V. Buldyrev and 
H. E. Stanley, Phys. Rev. E {\bf 71}, 061504 (2005).

\bibitem{Ca03}P. J. Camp, Phys. Rev. E. {\bf 68}, 061506 (2003).

\bibitem{Ca05}P. J. Camp, Phys. Rev. E. {\bf 71}, 031507 (2005).

\bibitem{He70} P.~C.~Hemmer and G.~Stell, Phys. Rev. Lett.
{\bf 24}, 1284 (1970); G.~Stell and P.~C.~Hemmer,
J.~Chem.~Phys. {\bf 56}, 4274 (1972); J.~M.~Kincaid,
G.~Stell, and C.~K.~Hall, {\it ibid.} {\bf 65}, 2161 (1976);
J.~M.~Kincaid, G.~Stell, and E.~Goldmark, {\it ibid.} {\bf
65}, 2172 (1976); C. K. Hall and G. Stell, Phys Rev. A
{\bf 7}, 1679 (1973); E. Velasco, L. Medeiros, G. Navascu\'{e}s,
P. C. Hemmer and G. Stell, Phys. Rev. Lett. {\bf 85}, 122 (2000);
 P. C. Hemmer, E. Velasco, L. Mederos, G. Navascu\'{e}s and G. Stell,
J. Chem. Phys. {\bf 114}, 2268 (2001). 

\bibitem{Ja98} E. A. Jagla, Phys. Rev. E {\bf 58}, 1478 (1998);
E. A. Jagla, J. Chem. Phys. {\bf 110}, 451 (1999); E. A. Jagla,
J. Chem. Phys. {\bf 111}, 8980 (1999); E. A. Jagla, Phys.
Rev. E {\bf
63}, 061501 (2001); E. A. Jagla, Phys. Rev. E {\bf 63}, 061509
(2001).

\bibitem{Wi02}N. B. Wilding and J. E. Magee, Phys. Rev. E
{\bf 66}, 031509 (2002).

\bibitem{Ku04}P. Kumar, S. V. Buldyrev, F. Sciortino, E. Zaccarelli and
H. E. Stanley, Phys. Rev. E {\bf 72}, 021501 (2005).

\bibitem{Xu05} L. Xu, P. Kumar, S. V. Buldyrev, S.-H. Chen, P. 
H. Poole, F. Sciortino, and H. E. Stanley, Proc. Natl. Acad.
Sci. U.S.A. {\bf 102}, 16558 (2005).

\bibitem{Zh05}Z. Yan, S. V. Buldyrev, N. Giovambattista, and H. E. Stanley,
Phys. Rev. Lett. {\bf 95}, 130604 (2005).

\bibitem{Zh06}Z. Yan, S. V. Buldyrev, N. Giovambattista, P. G. Debenedetti, and
H. E. Stanley, Phys. Rev. E {\bf 73}, 051204 (2006).

\bibitem{Wi06}H. M. Gibson and N. B. Wilding, 
Phys. Rev. E {\bf 73}, 061507 (2006).

\bibitem{Ol06}A. B. de Oliveira, P. A. Netz, T. Colla, and M. C. Barbosa,
J. Chem. Phys. {\bf 124}, 084505 (2006). 

\bibitem{Ch96} C. H. Cho, S. Singh and G. W. Robinson,
 Phys. Rev. Lett. {\bf 76}, 1651 (1996); Faraday
Discuss. {\bf 103}, 19 (1996); J. Chem. Phys. {\bf 107}, 7979 (1997).

\bibitem{Ch97a} C. H. Cho, S.~Singh, and G.~W. Robinson.  
Phys. Rev. Lett.  {\bf 79}, 180 (1997).

\bibitem{Ne04}Paulo A. Netz, J. Fernando Raymundi, Adriana Simone Camera   
and Marcia C. Barbosa, Physica A {\bf 342}, 48 (2004).

\bibitem{inspired} This cut-off was choosen inspired in the Refs. \cite{Zh05,Zh06}. An alternative
might have been the distance corresponding to the first or second minimum in $g.$ These different options do not affect the location of $\rho_{t-max}$ and
$\rho_{t-min}$.



\bibitem{Tr00} T. M. Truskett, S. Torquato, and P. G. Debenedetti, Phys. Rev. E {\bf 62}, 993 (2000).


\bibitem{Hu04}A. Huerta, G. G. Naumis, D. T. Wasan, D. J. Henderson, and A. D. Trokhymchuk, J. Chem Phys. {\bf 120}, 1506 (2004).

\bibitem{orders}In Refs. \cite{Er03,To00,Tr00,Hu04} the average of $Y_{lm}$ is taken over \emph{all} bonds
in the system. Therefore, no 'local order' concept exists. On the other hand, in Refs. \cite{Er01,Sh02} 
the average of $Y_{lm}$ is taken over the four nearest neighbors, quantifying a tetrahedral local
order for the system. The orientational order parameter of this work is based on the idea of local order 
for each particle, similar to Refs. \cite{Er01,Sh02}.

\bibitem{paper} We have tested the use of
a larger cutoff ($5.5\sigma$) and it does not affect the  results.

\bibitem{Nh85}W. G. Hoover, Phys. Rev. A {\bf 31}, 1695 (1985);
{\it ibid} {\bf 34}, 2499 (1986). 
 
\bibitem{clustering} This new peak may be interpreted as a signal of clustering.
In order to check this, some snapshots for $T^{*}=0.25$ and the range of densities
spanned by the Fig. \ref{cap:grs} were taken. These snapshots ( not shown) even displaying a non-randon distribution of particles, do no shown clearly 
any clustering.


\end{thebibliography}
\end{document}